\documentclass[11pt, a4paper]{article}

\usepackage{amsmath,amstext,amsbsy,amssymb}
\usepackage{fancyhdr,psfrag,latexsym}
\usepackage{epsfig}

\newcommand{\Ham}{\mathcal{H}}
\newcommand{\Hr}{\mathcal{H}_{\rho}}
\newcommand{\La}{\Lambda}
\newcommand{\PL}{P_{\Lambda}}

\begin{document}

\begin{titlepage}

\noindent
%\hspace*{11cm} Imperial/TP/95\\
\vspace*{1cm}
\begin{center} 
{\LARGE Quantum superposition principle and gravitational
    collapse: Scattering times for spherical shells}

\vspace{2cm}

M. Ambrus and P. H\'{a}j\'{\i}\v{c}ek \\
Institute for Theoretical Physics \\
University of Bern \\
Sidlerstrasse 5, CH-3012 Bern, Switzerland
\\
\vspace{1.5cm}

July 2005 \\ \vspace*{1cm}

\nopagebreak[4]

\begin{abstract}
  A quantum theory of spherically symmetric thin shells of null dust and their
  gravitational field is studied. In Nucl.\ Phys.\ {\bf 603} (2001) 515, it
  has been shown how superpositions of quantum states with different
  geometries can lead to a solution of the singularity problem and black hole
  information paradox: the shells bounce and re-expand and the evolution
  is unitary. The corresponding scattering times will be defined in the
  present paper. To this aim, a spherical mirror of radius $R_m$ is
  introduced. The classical formula for scattering times of the shell
  reflected from the mirror is extended to quantum theory. The scattering
  times and their spreads are calculated. They have a regular limit for
  $R_m\rightarrow 0$ and they reveal a resonance at $E_m = c^4R_m/2G$. Except
  for the resonance, they are roughly of the order of the time the light needs
  to cross the flat space distance between the observer and the mirror.  Some
  ideas are discussed of how the construction of the quantum theory could be
  changed so that the scattering times become considerably longer.

\end{abstract}

\end{center}

\end{titlepage}

\section{Introduction}
The most important and difficult problems of the relativistic theory of
gravitational collapse are the singularities (see, e.g., Ref.\ \cite{HE}) and
the black-hole information paradox (Ref.\ \cite{Gidd}). In fact, there are
some ideas around of how both the singularities and information losses in
gravitational collapse could be avoided. An example is a pioneer work by
Sacharov Ref.\ \cite{sach} in 60's assuming 1) that the equation of state for
very dense matter is $p = -\rho < 0$ so that its stress-energy tensor is
equivalent to a ``cosmological'' term, and 2) that the collapsed stuff will be
concentrated in a small, roughly stationary, very massive but everywhere
regular piece of matter with this equation of state. The space-time geometry
is mostly assumed to be classical and to obey classical Einstein's equations
everywhere: there seem to be no need for quantum gravity. The singularity
theorems of Ref.\ \cite{HE} are not applicable because of the large negative
pressure. For more detail and further developments see Ref.\ \cite{hay}.

Another proposal, which does not require exotic states of matter but exploits
the superposition principle of quantum theory instead, is described in Refs.\
\cite{HK}, \cite{H} and \cite{Honnef}. Let us quickly summarize what was shown
and what remained unclear there. In this way, we introduce and motivate what
will be done in the present paper.

The collapsing matter was represented by a spherically symmetric
self-gravitating thin shell of null dust. There were no other sources of
gravity and the space-time had a regular center. This was considered as a toy
model of black-hole creation. From the state of the shell at a space-like
3-surface that intersects the regular center and the shell, the gravitational
field can be completely determined by the constraints everywhere along the
3-surface. This was used to reduce the total action to variables that
described just the state of the shell (Ref.\ \cite{HK}).

In the quantum theory (Refs.\ \cite{H} and \cite{Honnef}), a self-adjoint
extension of the shell Hamiltonian (depending only on the shell state
variables) was chosen such that shell wave packets were formed by linear
combinations of in- and outgoing states and vanished at the center. In this
sense, the singularity was removed.

In principle, the quantum geometry around the shell ought to be determined by
the state of the shell. The problem was that the naive straightforward
calculation depended of the foliation along which the metric had to be
determined, and the resulting geometry itself was, unlike in the classical
theory, {\em strongly} dependent on it. This was interpreted in a more general
way as a failure of gauge-dependent methods in quantum gravity, which lead
in turn to search for some manifestly gauge-independent methods.

Still, the nature of the quantum horizon resulting in each quantum evolution
of the shell could be investigated in rough terms. It turned out that it was a
linear combination of states with black- and white-hole Schwarzschild
horizons. This was related to the fact that an in-going shell creates a
black-hole horizon, while an outgoing one creates a white-hole horizon outside
of it. Thus, the intriguing result that the shells with a sufficiently high
energy could cross their Schwarzschild radius in both directions could be
naturally explained.  This kind of quantum horizon was called {\em gray
  horizon} in Ref.\ \cite{H}.  It is a superposition of black- and white-hole
Schwarzschild horizons because the state of the shell is a superposition of
in- and outgoing motions. Now, the in- and outgoing states are related by the
time reversal, as are the white- and black-hole horizons.  Thus, the time
reversal seems to play some role here.

This role has to do with time reversal properties of Schwarzschild geometry,
which are rather subtle. If the Schwarzschild radial coordinate $R$ is larger
than Schwarzschild radius $2M$, $M$ being the mass parameter of the
Schwarzschild solution, the geometry is {\em locally} time-reversal invariant:
for any chosen point with $R>2M$, there is a time-reversal map that is an
isometry of the space-time and does not move the point. This is not true for
most points with $R \leq 2M$. It follows, e.g., that a contracting spherically
symmetric source of mass $M$ can create the same geometry outside as the same
source that expands, as long as the geometry is observed outside of $2M$. The
same is true for stationary axisymmetric black-hole space-times such as
Kerr's, only the time reversal is to be accompanied by an additional reversal
of the azimuthal coordinate.

This qualitative argument implies that quantum geometry, whatever it might be,
will differ strongly from the classical one if:
\begin{enumerate}
\item its (spherically symmetric) source is a superposition of in- and
  outgoing states,
\item the source gets under its Schwarzschild radius and
\item we measure the geometry under the Schwarzschild radius of the source.
\end{enumerate}
On the other hand, there seems to be no reason why the quantum geometry
outside the Schwarzschild radius had to differ much from the classical one if
the first two conditions hold.  Thus, our quantum superposition idea need not
contradict existence and observed properties of black-hole like objects,
because these properties are in any case theoretically calculated from the
classical geometry outside their horizons.

This is, however, only a speculation; one ought to calculate the quantum
geometry of the bouncing shell to prove the statement.  Such calculations are
very difficult. Problems begin already at the observation that we do not even
know how this quantum geometry is {\em defined}: it seems that no
gauge-invariant definition of it is known.  The effect of different gauge
choices is large in quantum gravity. Its full extent does not seem to be
adequately and sufficiently realized, although it is easy to assess. The gauge
fixing in general relativity can be understood as point to point
identification of manifolds with non-isometric geometries. As an example,
consider the average metric of, say, two different space-times with the same
topology. This notion is not well-defined (even in the classical physics),
unless the two manifolds are identified point by point. The resulting value of
the average depends on the identification so that, for some, it need not even
be a Lorentzian metric (see also Ref.\ \cite{Paris}).

The aim of the present work is more modest: We choose a particular measurable
property of the geometry and try to calculate it. It must be gauge invariant
because it is measurable but it need not contain information enough so that
all measurable properties of the geometry are determined by it. Which quantity
we choose? In Refs.\ \cite{H} and \cite{Honnef}, the quantum shell was shown to
contract from an asymptotically flat region, bounce and expand again into the
same asymptotically flat region. Hence, it must be possible for one and the
same observer in this region to meet the shell first in its contracting and
later in its expanding phase.  He can measure the time between the two events;
it is the so-called {\em scattering time} of the quantum shell. That is the
quantity we are going to calculate.

Although the scattering time contains only a small amount of information about
the quantum geometry, it is important. For example, if the scattering time is
too short, the existence of the gray horizon will also be short, and nothing
much can happen in its neighborhood to be observed as ``the black hole
properties''. If, however, the shell will come out only after many thousands
of years, or if its scattering time is even larger than the age of the
Universe, then the black hole object created by it would last long enough to
act in its neighborhood as a black hole and to be observed.  To calculate
this simple quantity has turned out to be still surprisingly difficult.  In
calculations based on a gauge fixing, a foliation of all space-times with
in-going and outgoing shells is to be prescribed. Depending on this choice, the
scattering time can take any real value!

This is not the only difficulty. The {\it time delay} (see Refs.\
\cite{Goldberger} and \cite{Martin}) often used in quantum scattering theory
to define the duration of a scattering process is infinite for long-ranged
potential like the Coulomb or the gravitational one. While there exists a
regularization of the time evolution dependent on the fixed central charge for
the Coulomb potential (Ref.\ \cite{Dollard}) that renders the time delay
finite (Ref.\ \cite{Bolle:1983xp}), it does not work in the gravitational case
because here the role of the 'central charge' is played by the energy of the
shell which is not fixed but depends on the state. Since a state-dependent
regularization does not make much sense, one has to abandon the hope to define
the time delay in an analogous way as it was possible in Coulomb scattering.

The problem is even worse: for example, the {\it sojourn time} (Ref.\
\cite{Goldberger}), which is finite for finite regions also in the case of
long-range potentials, cannot be properly defined. The definition of the
sojourn time is based on a time average, where the time integrated over is the
Minkowski time in flat space-time. In the curved shell space-time it is not
clear which 'time' should be chosen and an appropriate time coordinate could
depend on the shell's energy, turning it into an operator in the quantum
theory. Problems then arise with a sensible definition of the sojourn time
because of non-commuting operators. All these difficulties can be thought of
as a scattering-theory version of the so-called {\em problem of time} in
quantum gravity (cf.\ Refs.\ \cite{Kucharrev}, \cite{Isham}). That's why we
make a more modest approach by defining the scattering time already on the
classical level and turn it into an operator in a suitable quantum theory.

Any measurable classical geometrical property can be expressed in terms of
Dirac observables and ``quantized'' by choosing some factor ordering of the
corresponding operators. However, the scattering time is not such a property
because no classical shell bounces and re-expands but just disappears in the
black hole it creates.

The method that will be adopted in the present work (after quite a number of
different approaches have been attempted unsuccessfully) is as follows. First,
we shall modify the model by introducing a spherical mirror of radius $R_m$.
Then even a classical shell will be reflected by the mirror and there will be
a gauge-invariant classical formula for the scattering time. However, a
classical shell will be reflected to the asymptotic region from which it has
come only if its energy $E$ is smaller than the {\em critical energy}
\[
  E_m = \frac{R_mc^4}{2G}\ .
\]
$E_mc^{-2}$ is the mass whose Schwarzschild radius coincides with the radius
of the mirror. A shell with $E<E_m$ will not cross its own Schwarzschild radius
anywhere on its way to the mirror. Its classical scattering time will diverge
if its energy approaches $E_m$.

Second, it will turn out that the average scattering time in the quantum
theory does {\em not} diverge if the expected energy of the shell approaches
$E_m$, but has only a finite peak, as if there was a resonance. One can then
extend the quantum theory in an obvious way to cover energies larger than
$E_m$. Third, we take the limit $R_m \rightarrow 0$ in such an extended theory
and assume that the resulting theory gives the valid description of the
quantum collapse without the mirror.

The plan of the paper is as follows. In Sec.\ 2, all solutions with the
reflected shell are found. A complete set of Dirac observables is chosen and
symmetries in the space of solutions are listed. Then the classical formula
for the scattering time is derived. Sec.\ 3 summarizes relevant notions and
equations of the Hamiltonian formalism for null shells from Louko, Whiting and
Friedman paper \cite{LWF} (LWF). In Sec.\ 4, the two LWF actions, one for in-
and the other for the outgoing shells are unified and the action is modified
to include the mirror and shell reflections. The mirror is considered as a
formal boundary represented by some boundary conditions. This allows some
freedom of how the shell is reflected especially in the quantum theory.
Boundary conditions for the gravitational field and the shell at the mirror of
the classical version of the theory are chosen such that the desired solutions
result. Sec.\ 5 uses the method of Refs.\ \cite{HK} and \cite{HKou} to reduce
the action to the Dirac observables. In this way, the Poisson algebra of the
observables is determined. The observables are the energy $E$ of the shell and
its canonical conjugate $v$, which is the asymptotic advanced time of the
in-going shell.

This algebra forms a starting point for the construction of the quantum
mechanics in Sec.\ 6. We postulate a natural cut-off $E_o$ on energies that
may be used in the scattering states; it is the energy that would create a
horizon at the radius $R_o$ of the observer. In such a way, the energy is not
only positive but also bound from above. This makes it possible to find
self-adjoint extensions of $v$. The spectrum of $v$ is then discrete but as
dense as indistinguishable from a continuous one. The scattering times and the
times at which the observer meets the shell are turned to operators after the
classical formulae are extended for all scattering energies in Sec.\ 7. The
expected values and the spreads of these operators are calculated in Sec.\ 7.1
for eigenstates of $v$. They are all finite though the classical formula
has a singularity and they are independent of (the eigenvalue of) $v$.  Sec.\
7.2 studies the energy dependence of the scattering time and its spread using
simple box wave packets. Two kinds of phenomena are found. First, the
scattering times have a narrow peak at the critical energy $E_m$ (where the
classical formula has a singularity). This looks like a ``resonance''. Second,
if the energy increases beyond about 0.1 $E_o$, the scattering time reaches
another maximum and then begins to decrease eventually falling under zero near
$E_o$. This is due to the changes of geometry near the observer created by the
huge energy of the shell so that the scattering theory method ceases to be
applicable. Sec.\ 8 discusses our results and methods in some broader contexts.

This paper is based on Ref.\ \cite{ambrus}

\section{The model}
\begin{figure}[htbp]
 % \centering
% \hspace*{2.8cm}
  % \vspace{-0.2cm}

  \epsfig{figure=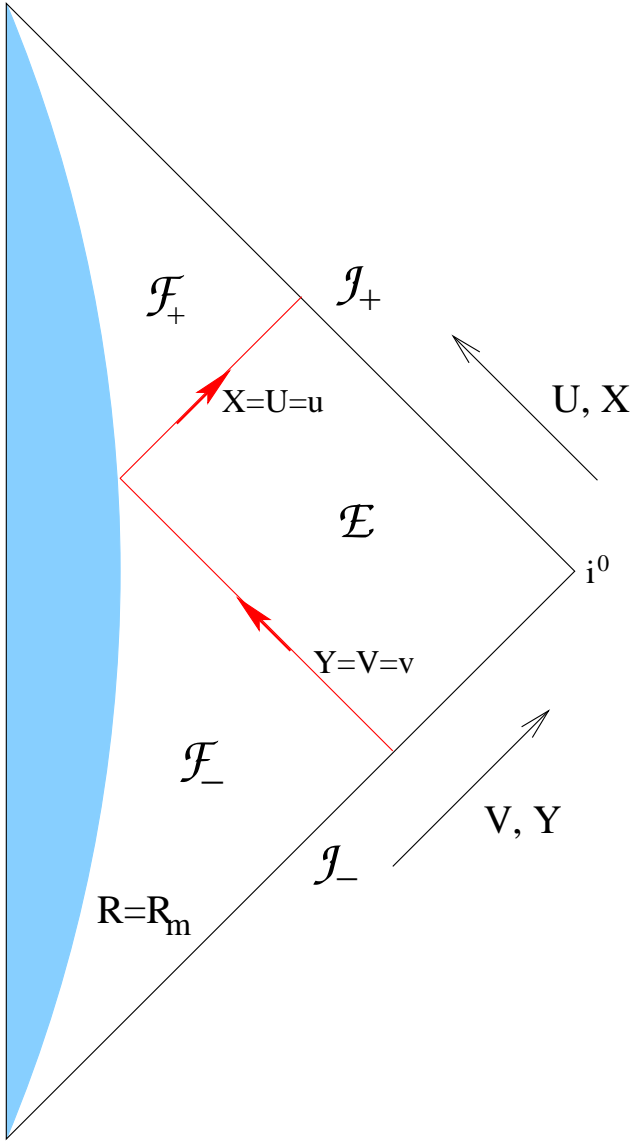}

  \caption[Diagram of the shell space-time $\bar{\mathcal{M}}$]{\it (Color
    online) Penrose-like diagram of the space-time $\bar{\mathcal{M}}$. The
    shaded region lies inside the mirror with the radius $R=R_m$. The in-going
    shell trajectory defined by $Y=V=v$ starts at past light-like infinity
    $\mathcal{I}_-$, becomes an outgoing shell trajectory $X=U=u$ at the
    mirror and ends up at future light-like infinity $\mathcal{I}_+$. The
    region outside the shell is denoted by $\mathcal{E}$, that inside the
    outgoing (in-going) shell by $\mathcal{F}_+$ ($\mathcal{F}_-$). The arrows
    show the directions in which the double-null coordinates $U$ and $V$ (or
    $X$ and $Y$) increase.  }
   \label{fig:shell mirror}
\end{figure}

We consider a self-gravitating, spherically symmetric and infinitesimally thin
shell such as in Ref.\ \cite{HK}. Unlike Ref.\ \cite{HK}, however, we
introduce a spherical ideal mirror of radius $R_m$ so that the shell and the
mirror are co-centric. The shell scatters at the mirror as depicted in Fig.\
\ref{fig:shell mirror}.

If such a shell of finite momentum is reflected at the mirror, the finite
change of its momentum must result within an infinitesimal time interval. This
requires the mirror to bear infinite force, an idealization similar to the
whole notion of a thin shell, and it must be understood only as a limiting
case of a regular system.

We shall view the mirror just as a boundary conditions at $R=R_m$ and
consider only the part of the space-time that lies outside of the mirror. The
conditions will be such that no energy and momentum can cross the boundary,
and that the mass of the mirror is zero. It will be specified more precisely
later. Logically, we define the solutions first, and only then infer the
corresponding boundary conditions from it.

In the present section, we construct and discuss the space of solutions for
the system that we have introduced in the previous paragraph. The space of
solutions for a single shell without the mirror is already well-known
(cf.~e.g.~Ref.\ \cite{HK}). Also the case of multiple shells has been
discussed in the literature, Ref.\ \cite{HKou}.  Any solution with a reflected
shell can be constructed by cutting and pasting together one solution with an
in-going and one with an outgoing shell. These can be taken over from Ref.\
\cite{HK}.

In Ref.\ \cite{HK}, the outgoing-shell space-time is described in the
double-null (DN) coordinates $U$ and $V$. The metric has everywhere the form
\begin{equation}\label{10.1}
  ds^2 = -A_+dUdV + R^2_+ d\Omega^2\ ,
\end{equation}
where $\Omega^2$ is the line element of a unite sphere. Outside the shell,
$-\infty < U < u$,
\begin{eqnarray}\label{11.1}
  A_+ &=& \frac{1}{\kappa(f_+)\exp(\kappa(f_+))}\ \frac{V-u}{4M_+}\
  \exp\left(\frac{V-U}{4M_+}\right)\ , \\
\label{11.2}
  R_+ &=& 2M_+\kappa(f_+)\ , \\
\label{11.3}
  f_+ &=& \left(\frac{V-u}{4M_+} - 1\right)\exp\left(\frac{V-U}{4M_+}\right)\ ,
\end{eqnarray}
where $\kappa(x)$ is the function inverse to $\kappa^{-1}(x) = (x-1)e^x$.
Metric (\ref{10.1}) describes Schwarzschild geometry with the mass parameter
$M_+$ related to the outgoing-shell energy $E_+$ by $M_+ = Gc^{-4}E_+$. Inside
the shell, $u < U < V$,
\begin{equation}\label{11.7}
  A_+ = 1\ ,\quad R_+ = \frac{V-U}{2}\ ;
\end{equation}
it is Minkowski space-time. Metric (\ref{10.1}) is continuous across the
shell, $U=u$. The parameters $M_+$ and $u$ have been chosen in Ref.\ \cite{HK}
as coordinates in the space of outgoing-shell solutions.

The in-going-shell space-time, given in the DN coordinates $X$ and $Y$, has the
metric
\begin{equation}\label{11.4}
  ds^2 = -A_-dXdY + R^2_-d\Omega^2\ .
\end{equation}

Outside the shell,
$v < Y < \infty$,
\begin{eqnarray}\label{11.5}
  A_- &=& \frac{1}{\kappa(f_-)\exp[\kappa(f_-)]}\ \frac{v-X}{4M_-}\
  \exp\left(\frac{Y-X}{4M_-}\right)\ , \\
\label{11.6}
  R_- &=& 2M_-\kappa(f_-)\ , \\
\label{12.1}
  f_- &=& \left(\frac{v-X}{4M_-} - 1\right)\exp\left(\frac{Y-X}{4M_-}\right)\ ,
\end{eqnarray}
$M_-$ being the Schwarzschild mass parameter related to the in-going-shell
energy $E_-$. Inside the shell  $X < Y < v$,
\[
  A_- = 1\ ,\quad R_- = \frac{Y-X}{2}\ .
\]
This is again flat space-time. The parameters $M_-$ and $v$ play the role of
coordinates in the corresponding space of solutions.

An important property is that the solution determined by the parameters $M_+$
and $u$ is isometric to that with $M_-$ and $v$ if $M_+ = M_-$. The
isometry is described by the relations
\begin{equation}\label{TR}
  U - u = v - Y\ ,\quad V - u = v - X
\end{equation}
and will be called {\em time reversal}.

\begin{figure}[htbp]
  \centering
  % \hspace*{0.7cm}
  
    \epsfig{figure=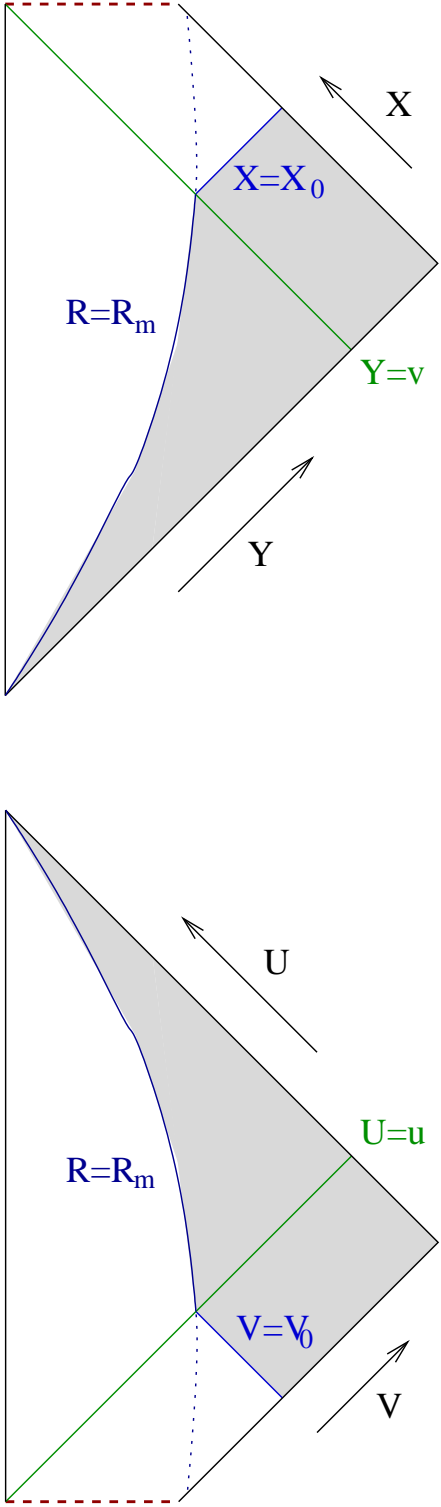}
  
    \caption[HK bild out] {\it (Color online) The diagram on the right hand
      side displays the Penrose diagram of the outgoing Ref.\ \cite{HK} shell
      space-time. The shell trajectory is given by $U=u$. The shaded region is
      cut out and glued with the corresponding shaded region taken from the
      in-going space-time on the left hand side. Here, the shell trajectory is
      given by $Y=v$. The rectangles outside of the shell are isometric.}
  \label{fig:HK beide}
\end{figure}

So much about the description of the solutions in Ref.\ \cite{HK}. Let us now
cut the outgoing-shell space-time along the curve $R=R_m$ inside the shell.
The coordinate $V$ of the intersection of this cut with the shell, $U=u$, is
\begin{equation}\label{12.3}
  V_0 = u + 2R_m\ .
\end{equation}
Our cut continues outside the shell along $V=V_0$. Finally, we throw away
everything inside the cut (see Fig.~\ref{fig:HK beide}). 

For the construction, the assumption is crucial that $R_+$ increases with
decreasing $U$ along the part of the cut that lies outside the shell, reaching
eventually $R_+=\infty$. This can only be true if $R_m$ is larger than the
Schwarzschild radius of the shell,
\begin{equation}\label{12.4}
  R_m > 2M_+\ ,
\end{equation}
or
\begin{equation}\label{12.5}
  E_+ < E_m\ ,
\end{equation}
the critical energy (defined in the Introduction).

Similarly, we cut the in-going-shell space-time at $R=R_m$ inside the shell
until the cut reaches the shell at 
\begin{equation}\label{13.1}
  X_0 = v - 2R_m\ ,
\end{equation}
and then proceed with the cut along $X=X_0$ outside the shell and throw
everything away that lies inside the cut (cf.~Fig.~\ref{fig:HK beide}). 
Let us further choose the parameters $M_-$ and $v$ as follows:
\begin{eqnarray}\label{13.2}
  M_- &=& M_+ = M\ , \\
\label{13.3}
  v &=& V_0\ .
\end{eqnarray}
This implies first that the outgoing null surface $X=X_0$ is diverging, $R_-$
increasing with $Y$ along it. Second, Eqs.\ (\ref{13.1}), (\ref{13.3}) and
(\ref{12.3}) imply 
\begin{equation}\label{13.4}
  v = u + 2R_m
\end{equation}
and 
\begin{equation}\label{13.5}
  X_0 = u\ .
\end{equation}
Also, the geometries of the two cut-out space-times are isometric in the
patches outside the shells, that is, respectively, in $U \in (-\infty,u)$, $V
\in (v,\infty)$ and $X \in (-\infty,u)$, $Y \in (v,\infty)$ and can be pasted
together there by the map
\begin{eqnarray}\label{14.1}
  X &=& v - 4M\kappa\left[\left(\frac{v-u}{4M} -
  1\right)\exp\left(\frac{v-U}{4M}\right)\right], \\ 
\label{14.2}
  Y &=& u +
  4M\ln\left[\frac{\kappa^{-1}\left(\frac{V-u}{4M}\right)}{\frac{v-u}{4M} -
  1}\right]\ , 
\end{eqnarray}
resulting in the region $\mathcal{E}$ of Fig.\ 1. One easily verifies that the
map is an isometry and that the corresponding boundaries cover each other.

In this way, we obtain solutions in which the outgoing shell crosses the
in-going one at $R = R_m$. All such solutions are described by just two
parameters, $M$ and $u$ (or $M$ and $v$, $u$ and $v$ being related by Eq.\ 
(\ref{13.4})). This defines the space of solutions of our system. Each
solution is given in two coordinate patches. This will not lead to any
problems later, and the fact that the coordinates of the patches coincide with
the coordinates chosen in Ref.\ \cite{HK} will enable us to use directly many
results of Ref.\ \cite{HK} without need for extra calculations.

In the space of solutions, we find the following symmetries. 

\noindent {\em The time shift}. This is the map 
\begin{equation}\label{timeshift1}
  U \mapsto U-\tau\ ,\quad  V \mapsto V-\tau\ ,\quad
  X \mapsto X-\tau\ ,\quad Y \mapsto Y-\tau\ , 
\end{equation}
for any real $\tau$. The solution obtained in this way from the solution with
parameters $u$, $v$, $M$ and $R_m$ corresponds to the change of the parameters
\begin{equation}\label{timeshift2}
  u \mapsto u+\tau\ ,\quad v \mapsto v+\tau\ , \quad
  M \mapsto M\ ,\quad R_m \mapsto R_m\ . 
\end{equation}

\noindent {\em The dilatation}. The map is defined by point shift
\[
  U \mapsto U/\xi\ ,\quad V \mapsto V/\xi\ , \quad
  X \mapsto X/\xi\ ,\quad Y \mapsto Y/\xi\ , 
\]
and metric deformation
\[
  g_{\mu\nu} \mapsto \xi^2g_{\mu\nu}\ .
\]
We obtain the solution with changed parameters
\[ 
  u \mapsto \xi u\ , \quad v \mapsto \xi v\ , \quad
  M \mapsto \xi M\ ,\quad R_m \mapsto \xi R_m\ . 
\]

The most interesting question for the present paper is how long it takes till
a shell returns to an observer at a fixed radius $R_o$: the so-called {\em
  scattering time} $s(R_o)$. The segment of the observer trajectory that
is bounded by the two intersections of the shell with it lies inside
each of the two patches. Let us calculate it in the coordinates $U$ and $V$.

The trajectory must satisfy the equation $R_+(U,V) = R_o$; Eqs.\
(\ref{11.2}) and (\ref{11.3}) give
\[
  U = V - 2R_o + 4M\ln\left(\frac{V-u-4M}{2R_o-4M}\right)\ .
\]
We also obtain from Eqs.\ (\ref{11.1}) and (\ref{11.2})
\[
  f_+ = \kappa^{-1}\left(\frac{R_o}{2M}\right)
\]
and 
\[
  A_+ = \left(1-\frac{2M}{R_o}\right)\ \frac{V-u}{V-u-4M}\ .
\]
The boundaries of the segment contain the crossing of the in-going shell at
$V=v$ and that of the outgoing one at $U=u$, where $V=u+2R_o = v-2R_m +2R_o$,
according to Eq.\ (\ref{13.4}). Then,
\[
  s(R_o) = \int_v^{v+2R_o-2R_m}dV\sqrt{A_+\frac{dU}{dV}}\ ,
\]
and we obtain easily
\begin{equation}\label{17.1}
  s(R_o) = \sqrt{1-\frac{2M}{R_o}}\left[2(R_o-R_m)
  +4M\ln\left(\frac{R_o - 2M}{R_m - 2M}\right)\right]\ .
\end{equation}
Here, the first factor transforms Schwarzschild time into the proper time of
the observer. The first term in the brackets is the flat-space-time scattering
time (the velocity of light is set to 1) and the second term is another
correction to it. Observe that this correction diverges if the energy of the
shell approaches the critical energy $E_m$.

Another gauge-invariant quantity is the proper time measured by the observer
when the shell is passing him, $s_+(R_o)$ for the outgoing and $s_-(R_o)$ for
the in-going shell. We have, of course
\begin{equation}\label{43.1}
  s(R_o) = s_+(R_o) - s_-(R_o)\ .
\end{equation}
The time $s_-(R_o)$ coincides with the flat-space-time inertial-system time $T$
of this event. Inside the in-going shell, this time is related to $X$ and $Y$
by
\[
  T = \frac{X+Y}{2}\ ;
\]
the shell runs along the curve $Y = v$ and the radius satisfies
\[
  R = \frac{-X+Y}{2}\ ,
\]
hence
\begin{equation}\label{43.2}
  s_-(R_o) = v - R_o\ .
\end{equation}
The other passing time, $s_+(R_o)$, can then be obtained by Eq.\ (\ref{43.1}).

\section{Canonical Action}
We are going to describe a canonical formalism for the dynamics of the
reflected shells. We start from LWF action \cite{LWF} and modify it so that it
describes both in- and outgoing shells; we also replace the fall-off
conditions at the internal infinity or the boundary conditions at the regular
center by the boundary conditions appropriate at the mirror.

Let us first summarize the relevant equations of the LWF paper. The space-time
geometry is given by the ADM metric
\begin{equation}
  ds^2 = -N^2 dt^2 + \La^2 (d\rho + N^{\rho} dt)^2 + R^2 d\Omega^2, 
\end{equation}
where $N, N^{\rho}, \La, R$ are functions of $t$ and $\rho$; $N, \La$ and $R$
are positive. $\La(\rho)$ and $R(\rho)$ are canonical coordinates of the
gravitational field while $P_\La(\rho)$ and $P_R(\rho)$ are their conjugate
momenta. The shell history is denoted by $\rho = r(t)$ and $p$ is the momentum
conjugate to $\rho$. A quantity $Q$ evaluated at the shell will be denoted by
$Q(r)$.  Derivatives with respect to $\rho$ are abbreviated by a prime, $Q'$,
those with respect to $t$ by an over-dot, $\dot{Q}$.

Including the null shell, the Hamiltonian bulk action reads
\begin{equation}
\label{eq:LWF}
  S_\eta = \int dt \left[ p\dot{r} + \int_{\mathbb{R}} d \rho \left( \PL
  \dot{\La} + P_R \dot{R} - N\Ham - N^\rho \Hr  \right)  \right], 
\end{equation}
where the constraint functions are given by
\begin{equation}
\label{eq:LWF Superhamiltonian}
  \Ham = \frac{\La \PL^2}{2R^2} - \frac{\PL P_R}{R} + \frac{RR''}{\La} -
  \frac{RR' \La'}{\La^2} + \frac{R'^2}{2\La} - \frac{\La}{2} + \frac{\eta
  p}{\La} \delta (\rho -r), 
\end{equation}
which is the so-called {\it super-Hamiltonian}, and 
\begin{equation}
\label{eq:LWF Supermomentum}
  \Hr = P_R R' - \PL' \La - p\delta (\rho -r)
\end{equation}
(the {\it super-momentum}) and where $N$, $N^\rho$ are Lagrange multipliers.
The variable $\eta$ takes on two values, $-1$ for the in- and $+1$ for the
outgoing shell and it holds that $\eta = \text{sgn}(p)$. Thus, LWF action is,
in fact, a set of two independent actions, each valid for one of the two
possible motions.  The functions $N, N^\rho, R, \La$ are to be smooth
functions of $\rho$ everywhere, except at the shell, where they are only
continuous and may have finite jumps in their first derivatives. Also the
conjugate momenta $P_R, \PL$ are smooth except at the shell, where they have
finite discontinuities.  The most singular contributions come from the
explicit matter delta-terms in the constraints and the implicit
delta-functions appearing in $R''$ and $\PL'$.

Variation of the action with respect to the canonical variables and the
Lagrange multipliers yields the dynamical, 
\begin{eqnarray}
\label{LWF eqs of motion1}
  \dot{\La} & = & N \left( \frac{\La \PL}{R^2} - \frac{P_R}{R} \right) +
  (N^\rho \La)', \\
\label{LWF eqs of motion2}
  \dot{R} & = & - \frac{N \PL}{R} + N^\rho R', \\
\label{LWF eqs of motion3}
  \dot{\PL} & = & \frac{N}{2} \left[ - \frac{\PL^2}{R^2} - \left(
  \frac{R'}{\La}  \right)^2 + 1 + \frac{2\eta p}{\La^2} \delta (\rho -r)
  \right] - \frac{N'RR'}{\La^2} + N^\rho \PL', \\
\label{LWF eqs of motion4}
  \dot{P}_R & = & N \left[ \frac{\La \PL^2}{R^3} - \frac{\PL P_R}{R^2} -
  \left( \frac{R'}{\La}  \right)'  \right] - \left( \frac{N'R}{\La} \right)' +
  (N^\rho P_R)', \\
\label{LWF eqs of motion5}
  \dot{r} & = & \frac{\eta N(r)}{\La (r)} - N^\rho (r), \\
\label{LWF eqs of motion6}
  \dot{p} & = & p \left( N^\rho - \frac{\eta N}{\La}  \right)' (r) , 
\end{eqnarray}
and the constraint equations 
\begin{equation}
  \label{eq:constraints}
  \Ham = 0, \quad \Hr = 0. 
\end{equation}

The possible occurrence of surface terms has not been taken into account yet,
but we will do it after having imposed the fall-off conditions on the metric
variables at the infinity $\rho \to \infty$, that have been given by Ref.\
\cite{LWF} and Kucha\v{r} \cite{Kuchar2}:
\begin{eqnarray}
\label{eq:LWF fall-off1}
  \La (t, \rho ) & \approx & 1 + \frac{M}{\rho} + O(|\rho|^{-1-\epsilon}),
  \\
\label{eq:LWF fall-off2}
  R (t, \rho ) & \approx & |\rho| + O(|\rho|^{-\epsilon}), \\
\label{eq:LWF fall-off3}
  \PL (t, \rho ) & \approx & O(|\rho|^{-\epsilon}), \\
\label{eq:LWF fall-off4}
  P_R (t, \rho ) & \approx & O(|\rho|^{-1-\epsilon}), \\
\label{eq:LWF fall-off5}
  N (t, \rho ) & \approx & N_{\infty} + O(|\rho|^{-\epsilon}), \\
\label{eq:LWF fall-off6}
  N^\rho (t, \rho ) & \approx & O(|\rho|^{-\epsilon}), 
\end{eqnarray}
where $M$ and $N_{\infty}$ are functions of $t$, and where $\epsilon \in
(0,1]$. For a solution, the function $M(t)$ becomes constant and coincides
with our parameter $M$. With these fall-off conditions the asymptotic region
($\rho \to \infty$) is asymptotically flat. $N_{\infty}$ is the rate at which
the asymptotic Minkowski time $T_{\infty}$ evolves with respect to the
coordinate time $t$. The variation of the term $\frac{NRR'\La'}{\La^2}$ in the
super-Hamiltonian constraint $-N\Ham$ with respect to $\La$ leads to the
non-vanishing surface term (cf.~Ref.\ \cite{Kuchar2})
\begin{eqnarray}
\label{eq:surface terms LWF}
  N_{\infty} \lim_{\rho \to \infty} \left( \frac{RR'}{\La^2} \delta \La
  \right) = N_{\infty} \delta M. 
\end{eqnarray}

The surface term (\ref{eq:surface terms LWF}) can be canceled by adding the
so-called ADM boundary term (see also Ref.\ \cite{Regge}) to the bulk action:
\begin{equation}
\label{eq:ADM term}
  S_{\infty} = - \int dt (N_\infty E_\infty)\ .
\end{equation}

\section{Removing $\eta$ and Introducing the Mirror}
Our action has to describe both in- and outgoing motions of the shell without
this options being predetermined by a chosen value of the variable $\eta$. This
is easy to arrange by replacing $\eta$ by $\text{sgn}(p)$. For example, in
Eq.\ (\ref{eq:LWF Superhamiltonian}) $\eta$ appears in the combination $\eta
p$, and we just write $|p|$. The origin of the absolute value here is simply
that it is to be the energy of the shell, and the energy $E$ of zero-rest-mass
particles and shells is $|p|$ instead of $\sqrt{p^2 + \mu^2}$. The range of
$p$ must be extended to $(\infty,\infty)$ and the sign of $p$ will follow
automatically from the equations of motions and the initial and boundary
conditions.

At the mirror, the geometry of the solutions is flat similarly as at the
infinity and we can choose the boundary conditions on the geometry there by
assuming that the foliation at the mirror is special in an analogous way to
that at the infinity.

We first suppose that the parameter $\rho$ assumes the fixed value $R_m$ at the
mirror,
\begin{equation}\label{19.1}
  \rho|_m = R_m\ ,
\end{equation}
and that the foliation is orthogonal to the mirror:
\begin{equation}\label{19.2}
  N^\rho|_m = 0\ .
\end{equation}
The lapse function can be left arbitrary,
\begin{equation}\label{19.3}
  N|_m = N_m(t) = \frac{dT}{dt}\ ,
\end{equation}
where $T$ is the Minkowski time at the mirror. Finally, we require that $\rho$
coincide with $R$ to the first order inclusively:
\begin{equation}\label{19.4}
  R'|_m = 1\ ,
\end{equation}
so that
\begin{equation}\label{19.5}
  \La|_m = 1\ .
\end{equation}
  
The modified LWF action reads:
\begin{equation}
  \label{eq:LWF mirror}
  S = \int dt \left[ p\dot{r} - N_\infty E_\infty + \int_{\rho_m}^\infty d
  \rho \left( \PL 
  \dot{\La} + P_R \dot{R} - N\Ham - N^\rho \Hr  \right)  \right],
\end{equation}
where the super-Hamiltonian is: 
\begin{equation}
  \label{eq:modif Superhamiltonian}
\Ham = \frac{\La \PL^2}{2R^2} - \frac{\PL P_R}{R} + \frac{RR''}{\La} -
  \frac{RR' \La'}{\La^2} + \frac{R'^2}{2\La} - \frac{\La}{2} + \frac{|p|}{\La}
  \delta (\rho -r),   
\end{equation}
The other terms are identical to those in the LWF action.  Varying the action
(\ref{eq:LWF mirror}) with respect to the canonical variables leads to the new
{\it surface term} $B_m$ from the mirror:
\begin{equation}
  \label{eq:B_m}
  B_m = \left[ \frac{NR}{\La} \delta R' - \frac{NRR'}{\La^2} \delta \La -
  \frac{N'R}{\La} \delta R + N^\rho P_R
  \delta R  - N^\rho \La \delta \PL  \right]_{\rho = \rho_m}. 
\end{equation}

Inserting Eqs.\ (\ref{19.1})--(\ref{19.5}) into Eq.~(\ref{eq:B_m}) yields that
the boundary term from the mirror vanishes:
\[
  B_m=0.
\]
From Eqs.~(\ref{19.1})--(\ref{19.5}) and Eq.~(\ref{LWF eqs of motion2}), it
also follows that the so-called {\it mass function} Ref.\ \cite{LWF},
\begin{equation}
  \label{eq:mass function}
  \mathbf{M} = \frac{R}{2} \left[ 1 - \left(\frac{R'}{\La}\right)^2 +
  \left(\frac{\PL}{R}\right)^2  \right].  
\end{equation} 
vanishes at the mirror, 
\begin{equation}
  \label{eq:M_m=0}
  \mathbf{M}_m=0\ ,
\end{equation}
so that the mass of the mirror is zero as required.

Even with vanishing surface terms $B_m$, the variation of the action 
(\ref{eq:LWF mirror}) does not lead to the correct equations of motion yet. An
additional boundary condition has to be imposed in order that the shell is
really reflected at the mirror and does not pass through it unhindered. To
incorporate the reflection, the total momentum of the shell at the mirror must
be zero: 
\begin{equation}
  \label{eq:p constant at mirror}
  \lim_{t\rightarrow t_m+}p(t) + \lim_{t\rightarrow t_m-}p(t) = 0\ ,
\end{equation}
where $t_m$ is the time of the intersection between the shell and the mirror.
This means also that the absolute value of the momentum of the shell does
not change, when the shell is reflected by the mirror, and the energy is
conserved.  

The action (\ref{eq:LWF mirror}) with the modified constraint (\ref{eq:modif
  Superhamiltonian}), fall-off conditions (\ref{eq:LWF
  fall-off1})--(\ref{eq:LWF fall-off6}), the boundary conditions
(\ref{19.1})--(\ref{19.5}) and (\ref{eq:p constant at mirror}), as well as the
requirements of continuity constitute a complete system determining the
evolution so that our space of solutions results. This can be seen as follows.

Outside the mirror, our equations of motion are equivalent to LWF equations
except possibly for the explicit requirement that sgn$(p)$ is constant, which
we have not imposed. However, this also follows from the LWF equations of
motion. As is shown in Ref.\ \cite{LWF}, these equations imply that the
solution around the shell coincides with Schwarzschild space-time with
different mass parameters $M_-$ and $M_+$. The difference $M_+ - M_-$ is
related to $|p|$ and to the choice of the radial parameter $\rho$ at the shell
via Eq.\ (A2a) of Ref.\ \cite{LWF}:
\begin{equation}\label{21.1}
  |p| = -R(r)\Delta R'(r)\ ,
\end{equation}
$\Delta$ meaning the jump across the shell. It follows that $|p|$ cannot
vanish, if $\rho$ satisfies the conditions of regularity, i.e., if everywhere
$|R'| > \epsilon$, where $\epsilon$ is some positive number. As for the sign
of $p$, Eq.\ (A2b) of Ref.\ \cite{LWF} reads
\begin{equation}\label{21.2}
  p = -\La(r)\Delta P_\La(r)\ .
\end{equation}
Since $p$ cannot vanish and is equal to a jump of quantities that must
evolve continuously to both sides of the shell, it cannot change sign along
the shell motion outside the mirror.

At the mirror, however, Eqs.\ (\ref{21.1}) and (\ref{21.2}) lose their
meaning because there is no inside of the shell to calculate the jump. There,
the evolution must be prescribed by hand, and this is done by Eq.\ (\ref{eq:p
  constant at mirror}). It is compatible with $|p|$ changing smoothly and
being non-zero, and so it entails that the sign of $p$ must change through
the reflection.

\section{Reduction}
\begin{figure}[htb]
 % \centering
  % \hspace*{0.7cm}
  \hspace{5cm}
    \epsfig{figure=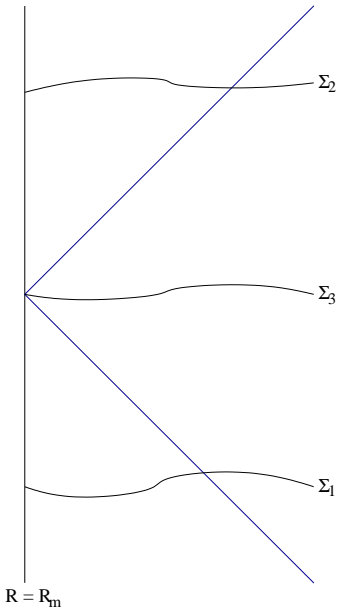}
  
    \caption[Embedding hyper-surfaces in the shell space-time] {\it (Color
      online) Schematic diagram of the trajectory of the shell bouncing at the
      mirror at the radius $R=R_m$. Three embedding hyper-surfaces $\Sigma$
      are drawn.  $\Sigma_1$ intersects the in-going shell, $\Sigma_2$ the
      outgoing one.  $\Sigma_3$ goes through the point where the shell hits
      the mirror.}
  \label{fig:Sigma mirror}
\end{figure}

We reduce the action (\ref{eq:LWF mirror}) using the same methods as in the
case without mirror. The shell's trajectory results from glueing together an
in- and an outgoing one at the mirror, in contrast to the system without
mirror, where the shell is either in- or outgoing. Thus, in our case, it
depends on where the embedding hyper-surface $\Sigma$ lies, which part of the
shell's trajectory it intersects. $\Sigma$ can even go through the point where
the shell hits the mirror. Fig.~\ref{fig:Sigma mirror} shows the three
possible cases for the embedding hyper-surface.

For the reduction of the action and for its expression in terms of the Dirac
observables, it is not important, which hyper-surface we choose. It will be
advantageous to choose a hyper-surface that intersects the outgoing shell, as
e.g.\ $\Sigma_2$ in Fig.\ \ref{fig:Sigma mirror} in the space-time
corresponding to the value $(M,u)$ of the Dirac observables. Such a
hyper-surface is related to a single point $Q$, say, at the constraint surface
$\mathcal C$ in the phase space of the system. All hyper-surfaces in a
neighborhood of $\Sigma_2$ in the space-time that intersect both the regular
center and the outgoing shell determine points in a neighborhood $\mathcal U$
of $Q$ at $\mathcal C$. If we carry out this construction for all space-times
corresponding to Dirac observables that lie in some neighborhood of the point
$(M,u)$ in ${\mathbf R}^2$, we fill up a neighborhood $\mathcal U$ of $Q$ at
$\mathcal C$.

Next, we choose the gauge $(U,V)$ so that we can describe each of the above
hypersurfaces by the embedding variables $U(\rho)$ and $V(\rho)$. In this way,
we have constructed coordinates $U(\rho)$, $V(\rho)$, $M$ and $u$ in the
neighborhood $\mathcal U$ in $\mathcal C$ (the momenta conjugate to $U(\rho)$
and $V(\rho)$ vanish at $\mathcal C$, see Ref.\ \cite{HK}). We can then, of
course, calculate the reduced action in these coordinates. However, we do not
need to do this explicitly, for exactly the same calculation has been done, in
effect, in Ref.\ \cite{HK}. That is the reason for all the choices above.

In Ref.\ \cite{HK}, it is first shown that the reduced action equals to its
Liouville form.  The Liouville form is expressed in terms of the embedding
variables and the shell variables. It is then shown that the Liouville form is
determined just by boundary terms at three boundaries: the regular center, the
shell and the infinity. The only change to be carried out here is that we have
the mirror instead of the regular center. It follows from the expressions in
HK that the Liouville form is still given by the boundary terms at the mirror,
the shell and the infinity. We have only to find the boundary term resulting
from the mirror.

This contribution from the mirror is given by Eq.\ (60) of Ref.\ \cite{HK} to
be
\begin{equation}\label{23.1}
  -(kdU + ldV)_{\rho=R_m} + \psi|_{\rho=R_m}d\rho_m\ ,
\end{equation}
where (cf.\ Eqs.\ (61) and (62) of Ref.\ \cite{HK})
\[
  k = \frac{R}{2}\frac{dR}{dU}\ln\left(-\frac{U'}{V'}\right)\ ,
\]
and 
\[
  l = \frac{R}{2}\frac{dR}{dV}\ln\left(-\frac{U'}{V'}\right)\ .
\]
The last term in Eq.\ (\ref{23.1}) vanishes because $\rho_m = R_m$ is
constant. Moreover, we have
\begin{equation}\label{23.2}
  \ln\left(-\frac{U'}{V'}\right) = 0\ ,
\end{equation}
because the foliation is orthogonal to the mirror. Indeed, if a radial curve
tangential to the mirror is given by $U = U(t)$, $V = V(t)$, $\vartheta =
\vartheta_0$ and $\varphi = \varphi_0$, then Eq.\ (\ref{11.7}) implies
\[
  V(t) - U(t) = 2R_m
\]
and the tangent vector $(\dot{U},\dot{V},\dot{\vartheta},\dot{\varphi})$,
therefore, satisfies
\begin{equation}\label{25.1}
  \dot{U} = \dot{V}\ ,\quad \dot{\vartheta} = \dot{\varphi} = 0\ .
\end{equation}
Let the foliation be given by $U = U(t,\rho)$ and $V = V(t,\rho)$. Then,
according to Eqs.\ (\ref{10.1}) and (\ref{11.7}), the foliation will be
orthogonal to the vector fulfilling (\ref{25.1}), if
\[
  -\dot{U}V' - \dot{V}U' = 0
\]
or $V' = -U'$, which immediately implies Eq.\ (\ref{23.2}).

It follows that the contribution from the mirror vanishes similarly as that
from the regular center did in Ref.\ \cite{HK}. Our Liouville form must,
therefore, coincide with that obtained in Ref.\ \cite{HK}, and so the reduced
action is
\begin{equation}\label{24.1}
  S = -M\dot{u}\ .
\end{equation}
This is equivalent to 
\begin{equation}\label{24.2}
  S = -M\dot{v}\ .
\end{equation}
because of  Eq.\ (\ref{13.4}). The same results can be obtained if the
calculation is carried out along the hyper-surfaces of the type $\Sigma_1$ or
$\Sigma_3$ of Fig.\ 2, but these have not been described explicitly in
Ref.\ \cite{HK}.

\section{Construction of the shell quantum mechanics}
The basic observables we have found in Sec.\ 5 are the energy $E$
and the asymptotic advanced time $v$ of the shell. They form a canonically
conjugate pair: $\{E,v \}=1$. In this section we turn these two observables
into self-adjoint operators on a suitable Hilbert space. We shall use units
for which $c=\hbar=1$. 

For quantization, the ranges of the classical observables are important. While
the values of $v$ take the whole real axis, $E$ must be positive and it has
been further limited in the classical theory to $E<E_m$ in order that a
scattering theory be applicable. We are going to extend this domain to higher
energies in quantum theory. However, even in quantum theory, it is impossible
that the shell energy is larger than
\[
  E_o = \frac{R_o}{2G}\ ,
\]
because the shells with $E>E_o$ create a black hole that includes the
observer. This is in contradiction with the basic assumption of the scattering
theory, namely that the observer is in the asymptotic region, where all
interactions are negligible and the geometry of space-time is practically
flat. As it will turn out, the scattering theory ceases to be applicable even
earlier. 

Thus, it is preferable to choose the interval $(0,E_o)$ as the range for $E$.
In this case, unlike to the half axis, a one-dimensional set of self-adjoint
operators $\hat{v}$ exists that satisfy the canonical commutation relations
with $\hat{E}$. They are all described in Ref.\ \cite{R-S}, PP. 141--142.

Let us choose the $E$-representation so that our Hilbert space can be
identified with $L^2(0,E_o)$, the space of square-integrable complex functions
$\psi(p)$ on the interval $(0,E_o)$, the action of the operator $\hat{E}$ being
\begin{equation}
  \label{45.1}
  \hat{E} \psi (p) = p \psi (p)\ ;
\end{equation}
the scalar product is
\begin{equation}
  \label{46.1}
  (\psi,\phi) = \int_0^{E_o}dp \psi^*(p)\phi(p)\ .
\end{equation}
Then, the self-adjoint operators $\hat{v}$ described in Ref.\ \cite{R-S} are
defined on the domain ${\mathcal D}_\theta$ of the absolutely continuous
functions satisfying the boundary conditions
\[
  \psi (E_o) = e^{i\theta}\psi (0)\ ,
\]
where $\theta\in[0,2\pi)$, by
\begin{equation}
  \label{46.2}
  \hat{v} \psi (p) = - i \partial_p \psi (p)\ .
\end{equation}
Clearly, on this domain, the canonical commutation relation are satisfied.

The eigenvalues and eigenfunctions of the thus defined operator
$\hat{v}$ are easily shown to be
\begin{equation}
  \label{eq:v theta EV}
  \hat{v} \phi_n (p,\theta) = \frac{2\pi}{E_o} \lambda \phi_n
  (p,\theta), \quad \lambda = n + \frac{\theta}{2\pi}, \quad n \in \mathbb{Z}
\end{equation}
and
\begin{equation}
  \label{eq:v theta EF}
  \phi_n (p) = \frac{1}{\sqrt{E_o}} e^{2i\pi\lambda \frac{p}{E_o}},
  \quad (\phi_m, \phi_n) = \delta_{mn}.
\end{equation}
The phase $\theta$ appears in the eigenfunctions $\phi_n$, so each measurement
of the operator $\hat{v}$ will depend on it. We have to choose a
particular value of $\theta$. The choice $\theta=0$ corresponds to the
periodical boundary condition often used for the momentum operator in a
box. This is surely the simplest choice.  
With this choice the eigenvalues and -functions of $\hat{v}$
are given by
\begin{equation}
  \label{eq: v EV}
  v_n = \frac{2\pi n}{E_o}, \quad n \in \mathbb{Z}, 
\end{equation}
and
\begin{equation}
  \label{eq:v EF}
  |v_n\rangle = \phi_n (p) = \frac{1}{\sqrt{E_o}} e^{2i\pi n \frac{p}{E_o}} =
  \frac{1}{\sqrt{E_o}} e^{i v_n p}.
\end{equation}

The spectrum of $\hat{v}$ is discrete. This is the price we have to pay for
making the operator self-adjoint on a finite interval. The discreteness of the
spectrum does not seem to correspond to a realistic physical
situation, where the asymptotic advanced time takes values from a continuous
spectrum. But the situation is not so bad as it seems if we look at the
distance between two neighboring eigenvalues, $d=v_{n+1}-v_n$: The distance
\begin{equation}
  \label{eq:distance of EV}
  d = \frac{2\pi}{E_o} = \frac{2\pi\hbar c}{E_o}
\end{equation}
becomes very small, $d \ll 1$, so that the spectrum can be considered 'almost
continuous'. Indeed, $E_o \approx 100\ M_{\text{Earth}}$ for $R_o \approx 1\ 
m$, and $d \approx 10^{-66}\ m$.

The eigenstates of $\hat{v}$ define the Fourier transform from $v$- to
$p$-representation  
\begin{equation}
  \label{eq:FT v to p}
  \psi(p) = \frac{1}{\sqrt{E_o}} \sum_{n=-\infty}^{\infty} e^{2i\pi  n
  \frac{p}{E_o}} \tilde{\psi} (n)
\end{equation}
and its inverse, 
\begin{equation}
  \label{eq:FT p to v}
  \tilde{\psi} (n) = \frac{1}{\sqrt{E_o}} \int_0^{E_o} dp \: e^{-2i\pi  n
  \frac{p}{E_o}} \psi(p).
\end{equation}

The final step in constructing a quantum theory is a choice of dynamics. In
our case, there is a time-translation symmetry given by Eqs.\ 
(\ref{timeshift1}); the corresponding change of Dirac observables described by
Eqs.\ (\ref{timeshift2}) is canonically generated by the energy $E$. Hence,
the most natural choice of Hamiltonian is $\hat{H} = \hat{E}$ (see
Ref.\ \cite{Honnef}). 

Let us work in Schr\"{o}dinger picture so that the wave function acquires the
time dependence
\[
  \psi(p)e^{-ipt}\ .
\]
This, of course, makes expected values of $\hat{v}$ time dependent. The
interpretation of this time dependence is, as explained in Ref.\
\cite{Honnef}, that a time-shifted observer will see a different value of $v$
in the same state.  For example,
\[
  \int_0^{E_o}dp\,\psi^*_n(p)e^{ipt}\hat{v}\left[\psi_n(p)e^{-ipt}\right] =
  v_n - t\ ,
\]
which is in agreement with Eqs.\ (\ref{timeshift2}) for $\tau = -t$, that is,
for an observer shifted by the amount $+t$ of time. There is no contradiction
in the claim that the expected value of $\hat{v}$ in a state is not equal to
some eigenvalue of $\hat{v}$.

\section{The scattering times}
In this section, we construct operators from the classical observables
$s_\pm(R_o)$ and $s(R_o)$ for the whole range of the energy $E \in
(0,E_o)$ and show that their expected values and spreads are finite in
reasonable states.

We define in $p$-representation:
\begin{equation}\label{48.1}
  \hat{s}_-(R_o) = -i\frac{\partial}{\partial p} - R_o
\end{equation}
and 
\begin{equation}\label{48.2}
  \hat{s}(R_o) = \sqrt{1-\frac{2Gp}{R_o}}\left[2(R_o-R_m) +
  4Gp\ln\left|\frac{1-\frac{2Gp}{R_o}}{\frac{R_m}{R_o}-
  \frac{2Gp}{R_o}}\right|\right]\ . 
\end{equation}
The multiplicative operator $\hat{s}(R_o)$ is indeed well-defined
on all continuous functions $\psi(p)$ because the result of its action on such
functions is square integrable.

Observe also that we can write Eq.\ (\ref{48.2}) in a scale-invariant form
\begin{equation}\label{48.3}
  \frac{\hat{s}(R_o)}{R_o} = 2\sqrt{1-q}\left(1 - \rho +
  q\ln\left|\frac{1-q}{\rho-q}\right|\right)\ ,
\end{equation}
where
\[
  \rho = \frac{R_m}{R_o} < 1
\]
and 
\[
  q = \frac{p}{E_o} \le 1
\]
are dimension-free quantities. The existence of such a formula is related to
the dilatation symmetry of the classical theory. On the other hand, Eq.\
(\ref{48.1}) cannot be written in such a form and we have only
\[
  \frac{\hat{s}_-(R_o)}{R_o} = -i\frac{2G}{R_o^2}\ \frac{\partial}{\partial q}
  - 1\ ,
\]
because the canonical commutation rules break the dilatation symmetry.

\subsection{Eigenstates of $\hat{v}$}
The calculation of $\langle v_n|\hat{s}(R_o)|v_n\rangle$ or 
$\langle v_n|[\hat{s}(R_o)]^2|v_n\rangle$ is straightforward but
tedious. All other expected values and spreads are, however, easily
expressible in terms of these two. For example, we have
\begin{multline}\label{eq:mean s_+^2}
  \langle v_n|[\hat{s}_-(R_o) + \hat{s}(R_o)]^2|v_n\rangle \\
  =\langle v_n|[\hat{s}_-(R_o)]^2|v_n\rangle + \langle
  v_n|[\hat{s}(R_o)]^2|v_n\rangle + 
  2|\sum_m\langle v_n|\hat{s}(R_o)|v_m\rangle  
  \langle v_m|\hat{s}_-(R_o)|v_n\rangle| \\
  = \langle v_n|[\hat{s}_-(R_o)]^2|v_n\rangle + \langle
  v_n|[\hat{s}(R_o)]^2|v_n\rangle  
  + 2\langle v_n|\hat{s}(R_o)|v_n\rangle \langle
  v_n|\hat{s}_-(R_o)|v_n\rangle 
\end{multline}
because the operator $\hat{s}_-(R_o)$ is diagonal in the basis $|v_n\rangle$.

Our method of calculation will be based on the formula
\begin{equation}\label{50.1}
  \int dq\,X(q)\ln|q-\xi| = [Y(q)-Y(\xi)]\ln|q-\xi| - \int dq\
  \frac{Y(q)-Y(\xi)}{q-\xi}\ ,
\end{equation}
where $Y(q)$ is any primitive function to $X(q)$,
\[
  Y(q) = \int dq\,X(q)\ .
\]
(The integration per partes is just performed here with the choice of
the primitive function that makes the resulting formula manifestly regular.)

Using Eqs.\ (\ref{eq:v EF}) and (\ref{48.3}), we obtain
\begin{multline}\label{50.2}
  \frac{1}{R_o}\langle v_n|\hat{s}(R_o)|v_n\rangle
  = 2\int_0^1dq\sqrt{1-q}\left(1-\rho + q\ln|q-1| - q\ln|q-\rho|\right) \\
  = -\frac{4}{3}(1-\rho)(1-q)^{3/2}|^1_0 + 2\int_0^1dq\,q\sqrt{1-q}\ln|q-1| 
  -2\int_0^1dq\,q\sqrt{1-q}\ln|q-\rho|\ .
\end{multline}
Now,
\[
  \int dq\,q\sqrt{1-q} = -\frac{2}{15}(2+q-3q^2)\sqrt{1-q}\ ,
\]
and we obtain from formula (\ref{50.1}), first,
\begin{multline*}
  \int_0^1dq\,q\sqrt{1-q}\ln|q-1| \\
  = -\frac{2}{15}(2+q-3q^2)\sqrt{1-q}\ln|q-1||^1_0 + \frac{2}{15}\int_0^1dq\
  \frac{(2+q-3q^2)\sqrt{1-q}}{q-1} \\
  = -\frac{2}{15}\int_0^1dq\,\sqrt{1-q}(2+3q) = -\left(\frac{8}{15}\right)^2
\end{multline*}
and, second,
\begin{multline*}
  \int_0^1dq\,q\sqrt{1-q}\ln|q-\rho| \\
  = -\frac{2}{15}\left[(2+q-3q^2)\sqrt{1-q} -
  (2+\rho-3\rho^2)\sqrt{1-\rho}\right]\ln|q-\rho||^1_0 \\
  +\frac{2}{15}\int_0^1dq\
  \frac{(2+q-3q^2)\sqrt{1-q} - (2+\rho-3\rho^2)\sqrt{1-\rho}}{q-\rho} \\
  = -\frac{4\times 31}{15^2} - \frac{8}{15}\rho +
  \frac{2}{15}[2-(2+\rho-3\rho^2)\sqrt{1-\rho}]\ln\rho +
  \frac{4}{15}(2+\rho-3\rho^2)\sqrt{1-\rho}\ln(1+\sqrt{1-\rho}) \ .  
\end{multline*}

In order to see, what this complicated formula means, we expand it in powers of
$\rho$ and $\ln\rho$, neglecting all terms with $\rho^2$, $\rho^2\ln\rho$ and
higher. The result is independent of $\rho$!
\begin{equation}\label{51.1}
  \frac{1}{R_o}\langle v_n|\hat{s}(R_o)|v_n\rangle = \frac{4}{15}(7
  - 4\ln 2)\ .
\end{equation}

Observe that the corresponding classical result for the flat space-time is
$2(1-\rho)$ and so the quantum scattering time is shorter that the classical
flat space-time one if $\rho < (1 + 8\ln 2)/15$, which is surely the case for
all $\rho\ll 1$. This is due to the broad spread of energy in the state
$|v_n\rangle$ and the factor $\sqrt{1-q}$ in the integral because $q$ runs up
to 1 in the quantum theory: the massive shells ``shorten'' the proper time
interval measured by the external observer.  The independence of the formula
on $v_n$ is clearly due to the time-translation invariance of the model.

The expected value of the square of the scattering time operator in the state
$|v_n \rangle$ is given by 
\begin{multline}
  \label{eq:DT2}
  \frac{1}{R_o^2} \langle v_n|(\hat{s}(R_o))^2|v_n\rangle =
  \frac{4}{3} - \frac{5}{3}\rho - \frac{2}{3}\rho^2+\rho^3+
  \left(\frac{2}{3}-\frac{11}{3}\rho+2\rho^2+\rho^3\right)  \rho \ln \rho \\ +
  \left(-1 +2 \rho^2 - \rho^4 \right)  \ln (1-\rho) + \left( \frac{4}{3} -
  \rho \right) \rho^3 \ln^2 \rho  +
  \frac{2}{3}(1 -4\rho^3 + 3\rho^4)\left(\frac{\pi^2}{3} - \text{dilog \,}
  \rho \right)
\end{multline}
where the dilogarithmic function is defined by
\begin{equation*}
  \text{dilog \,} x := \int dx \, \frac{\ln x}{1-x}.
\end{equation*}
Expansion in $\rho$ yields that also the squared scattering time operator does
not depend on $\rho$ up to first order:
\begin{equation}
  \label{eq:DT2 approx}
  \frac{1}{R_o^2} \langle v_n|(\hat{s}(R_o))^2|v_n\rangle \approx
  \frac{4}{3} + \frac{1}{9} \pi^2. 
\end{equation}
Here we have used the expansion of the dilogarithm,
\begin{equation*}
  \text{dilog \,} \rho \approx \frac{\pi^2}{6}
  - \rho +\rho \ln \rho + \mathcal{O}(\rho^2,\rho^2\ln\rho). 
\end{equation*}

The mean value of the operator $\hat{s}_- (R_o)$ is simply obtained:
\begin{equation}
  \label{eq:mean s_-}
  \langle v_n|\hat{s}_-(R_o)|v_n\rangle = - R_o + \frac{2\pi n}{E_o}.
\end{equation}
That of its square reads
\begin{equation}
  \label{eq:mean s_-^2}
  \langle v_n|\hat{s}^2_-(R_o)|v_n\rangle = R_o^2 - \frac{4\pi n R_o}{E_o} +
  \frac{4\pi^2 n^2}{E_o^2}, 
\end{equation}
thus the spread vanishes, as expected:
\begin{equation}
  \label{eq:Delta s_-}
  (\Delta s_-)_n = 0.
\end{equation}

The expected value of the operator $\hat{s}_+$ is easily found by using the
formulae (\ref{50.2}) and (\ref{eq:mean s_-}). Its square can be obtained by
using Eqs.~(\ref{eq:mean s_+^2}), (\ref{eq:DT2}), (\ref{eq:mean s_-}) and
(\ref{eq:mean s_-^2}). Expanding in $\rho$ as above, one finds the spread of
$\hat{s}_+$ to be independent on $\rho$ and on the state $|v_n \rangle$:
\begin{equation}
  \label{eq:spread s_+}
  \frac{\Delta s_+ (R_o)}{R_o} \approx \sqrt{\frac{4}{3}+\frac{\pi^2}{9} -
  \frac{4}{15}(7-4\ln 2)} \approx 1.1413.
\end{equation}
We observe that everything has a well-defined limit as $\rho\rightarrow 0$.

\subsection{Energy dependence of the scattering time}
In this subsection, we are going to study the behavior of the scattering
times with energy. To that aim, we have to work with suitable wave packets
$\phi(p)$. For example, to calculate $\langle \phi|\hat{s}(R_o)|\phi\rangle$
and $\langle \phi|[\hat{s}(R_o)]^2|\phi\rangle$, one can take the so-called
box wave packets,
\[
  \phi(p) = 0 \quad \forall p\in
  (0,E_o(\bar{q}-w/2))\cup(E_o(\bar{q}+w/2),E_o)
\]
and
\[
  \phi(p) = \frac{1}{\sqrt{E_ow}} \quad \forall
  p\in(E_o[\bar{q}-w/2],E_o[\bar{q}+w/2])\ ,
\]
where $E_o\bar{q}$ is the mean energy and $E_ow$ the width of the packet,
$\bar{q}$ and $w$ being the corresponding dimension-free quantities.
Operators containing $\hat{v}$ or $\hat{v}^2$ may then have diverging expected
values. However, the box wave packets are completely sufficient and perfectly
suitable for the study of the energy dependence of the scattering time and its
spread.

Analogously to the preceding subsection the expected values can be written in
terms of the dimension-free quantities:
\begin{equation}
  \label{eq:s bar box}
  \frac{1}{R_o}\langle \phi |\hat{s}(R_o)| \phi \rangle = \frac{1}{w}
  \int_{\bar{q}-w/2}^{\bar{q}+w/2} dq \, F(\rho,q) 
\end{equation}
and
\begin{equation}
  \label{eq:s2 bar box}
  \frac{1}{R_o^2}\langle \phi |[\hat{s}(R_o)]^2| \phi \rangle =
  \frac{1}{w} \int_{\bar{q}-w/2}^{\bar{q}+w/2} dq \, F^2(\rho,q) \ ,  
\end{equation}
where we have used the abbreviation
\[
  F(\rho,q) := 2\sqrt{1-q}\left(1 - \rho +
  q\ln\left|\frac{1-q}{\rho-q}\right|\right)\ .
\]
The spread is
\begin{equation}
  \label{81.1}
  \frac{\Delta\hat{s}(R_o)}{R_o} = \sqrt{\frac{1}{R_o^2}\langle \phi
  |[\hat{s}(R_o)]^2| \phi \rangle - \frac{1}{R_o^2}[\langle \phi
  |\hat{s}(R_o)| \phi \rangle]^2}\ . 
\end{equation}
The integrands are the same as in the case of the $v$-eigenstates and the
integrals can be computed in a completely analogous manner. Since they are
rather unwieldy, we will not write down the results explicitly but rather
derive important properties of the expected values using suitable
approximations.

In the interval
\[
  0 < q \ll \rho\ ,
\]
we can expand $F$ in powers of $q$:
\[
  F(\rho,q) \approx 2(1-\rho) - (2\ln\rho + 1 - \rho)q + \dots\ .
\]
Integrating term by term, we obtain
\[
  \frac{1}{R_o}\langle \phi |\hat{s}(R_o)| \phi \rangle \approx 2(1-\rho) -
  (2\ln\rho + 1 - \rho)\bar{q}\ ,
\]
and the leading term is the flat space-time value as expected. Numerical study
of the function $F$ shows that the expected value (\ref{eq:s bar box}) is
increasing in the whole interval $0 < \bar{q} < \rho$.

At $\bar{q}=\rho$, the integral in Eq.\ (\ref{eq:s bar box}) can be written as
follows 
\[
  \frac{1}{w}\int_{\rho-w/2}^{\rho+w/2}dq\ F(\rho,q) 
  = \frac{1}{w}\int_{\rho-w/2}^{\rho+w/2}dq\ a(q) -
  \frac{1}{w}\int_{\rho-w/2}^{\rho+w/2}dq\ b(q)\ln|q-\rho|\ , 
\]
where
\[
  a(q) := 2\sqrt{1-q}\ [1 - \rho + q\ln(1-q)]
\]
and 
\[
  b(q) := 2q\sqrt{1-q}
\]
are smooth functions in a neighborhood of $q=\rho$. We are going to expand
the integrals in powers of $w$ and of $\ln(w/2)$. We observe first that
\begin{equation}\label{91.1}
  \frac{1}{w}\int_{\rho-w/2}^{\rho+w/2}dq\ a(q) = a(\rho) + O(w)
\end{equation}
for any $C^1$ function $a(q)$. Second, we can use the trick of the foregoing
subsection: 
\begin{multline*}
  \frac{1}{w}\int_{\rho-w/2}^{\rho+w/2}dq\ b(q)\ln|q-\rho| \\
  = \frac{1}{w}\left\{[b_1(q) -
  b_1(\rho)]\ln|q-\rho|\right\}_{\rho-w/2}^{\rho+w/2} -
  \frac{1}{w}\int_{\rho-w/2}^{\rho+w/2}dq\ 
  \frac{b_1(q) - b_1(\rho)}{q-\rho}\ ,   
\end{multline*}
where $b'_1(q) = b(q)$ so that
\[
  \lim_{q\rightarrow\rho}\frac{b_1(q) - b_1(\rho)}{q-\rho} = b(\rho)\ .
\]
Hence, if $b(q)$ is any smooth function,
\begin{equation}\label{91.2}
  \frac{1}{w}\int_{\rho-w/2}^{\rho+w/2}dq\ b(q)\ln|q-\rho| = -b(\rho) +
  b(\rho)\ln(w/2) + O[w\ln(w/2)]\ .
\end{equation}
Collecting all results, we obtain
\begin{equation}\label{91.3}
  \frac{1}{R_o}\langle \phi |\hat{s}(R_o)| \phi \rangle  = 2\sqrt{1-\rho}\ [1
  + \rho\ln(1-\rho) - \rho\ln(w/2)] + O[w\ln(w/2)]\ .
\end{equation}
There is a sharp peak at $\bar{q}=\rho$ that grows like $-\ln\frac{w}{2}$.
From this we infer that the scattering time displays a kind of {\it resonance}
phenomenon near the critical energy, $\bar{q} \approx \rho$. The resonance
gets more distinct when the packet becomes narrower.

The integral in Eq.\ (\ref{eq:s2 bar box}) can be dealt with in an analogous
way. First,
\begin{multline*}
  \frac{1}{w}\int_{\rho-w/2}^{\rho+w/2}dq\ F^2(\rho,q) =
  \frac{1}{w}\int_{\rho-w/2}^{\rho+w/2}dq\ \bar{a}(q) \\
  -\frac{1}{w}\int_{\rho-w/2}^{\rho+w/2}dq\ \bar{b}(q)\ln|q-\rho| +
  \frac{1}{w}\int_{\rho-w/2}^{\rho+w/2}dq\ \bar{c}(q)\ln^2|q-\rho|\ , 
\end{multline*}
where
\begin{eqnarray*}
  \bar{a}(q) &:=& 4(1-q)[1-\rho+q\ln(1-q)]^2\ , \\
  \bar{b}(q) &:=& 8(1-q)q[1-\rho+q\ln(1-q)]
\end{eqnarray*}
and
\[
  \bar{c}(q) := 4(1-q)q^2
\]
are smooth functions in a neighborhood of $q=\rho$. Thus, for the first two
integrals, we can use the formulae (\ref{91.1}) and (\ref{91.2}). The third
integral can be written as follows
\begin{multline*}
  \frac{1}{w}\int_{\rho-w/2}^{\rho+w/2}dq\ \bar{c}(q)\ln^2|q-\rho| =
  \frac{1}{w}\left\{[\bar{c}_1(q) -
  \bar{c}_1(\rho)]\ln^2|q-\rho|\right\}_{\rho-w/2}^{\rho+w/2} \\
  -\frac{1}{w}\int_{\rho-w/2}^{\rho+w/2}dq\ 2\frac{\bar{c}_1(q) -
  \bar{c}_1(\rho)}{q-\rho}\ln|q-\rho|\ , 
\end{multline*}
where $\bar{c}'_1(q) = \bar{c}(q)$ and the function 
\[
  \bar{c}_2(q) := 2\frac{\bar{c}_1(q) - \bar{c}_1(\rho)}{q-\rho}
\]
is smooth in a neighborhood of $q=\rho$ with $\bar{c}_2(\rho) =
\bar{c}(\rho)$. Again, we use Eq.\ (\ref{91.2}) and obtain
\begin{multline}\label{93.1}
  \frac{1}{R^2_o}\langle \phi |(\hat{s}(R_o))^2| \phi \rangle  =
  4(1-\rho)\rho\ [\rho-4\ln(1-\rho)] \\
  +4(1-\rho)[1 + \rho\ln(1-\rho) - \rho\ln(w/2)]^2 + O(w\ln^2(w/2))\ .
\end{multline}
It follows that the spread of the scattering time at $\bar{q} = \rho$ attains
a regular limit for $w \rightarrow 0$ and in this sense is relatively weakly
dependent on $w$:
\begin{equation}\label{93.2}
  \frac{\Delta\hat{s}(R_o)}{R_o} = 2\sqrt{(1-\rho)\rho\ [\rho-4\ln(1-\rho)]}
  + O(w\ln^2(w/2))\ . 
\end{equation}

Finally, in the interval
\[
  \rho < q < 1\ ,
\]
numerical analysis shows that $F$ first steeply falls to a minimum, then
slowly increases reaching its second maximum if $\rho$ is smaller than about
0.1, and then falls again to negative values near $q=1$. Near its local
maximum, $F$ is of course slowly changing and the scattering time is in a very
good approximation equal to the value of $F$ there (for small $w$).  Expanding
$F$ and $\partial F/\partial q$ in powers or $\rho$, we find that the
position and the value of the second maximum depends only weakly on $\rho$; the
corrections to results obtained for $\rho = 0$ are of the second order in
$\rho$. Thus, we obtain for the position $q_{\text{M}}(\rho)$ of the second
maximum
\[
  q_{\text{M}}(\rho) = q_{\text{M}}(0) + O(\rho^2)\ ,
\]
where $q_{\text{M}}(0)$ is the larger solution to the equation $\partial
F/\partial q(0,q) = 0$, which reads
\[
  \ln\frac{1-q}{q} = \frac{3}{2-3q}\ ,
\]
and the value $F_{\text{M}}(\rho)$ of the second maximum is 
\[
\frac{F_{\text{M}}(\rho)}{R_o} = \frac{F_{\text{M}}(0)}{R_o} + O(\rho^2)\ ,
\]
where $F_{\text{M}}(0)$ is the value of the second maximum of $F(0,q)$, or
\[
  \frac{F_{\text{M}}(0)}{R_o} =
  \frac{4\sqrt{1-q_{\text{M}}(0)}}{2-3q_{\text{M}}(0)}\ .
\]
Numerical calculations yield
\[
  q_{\text{M}}(0) = .133071
\]
and 
\[
  \frac{F_{\text{M}}(0)}{R_o} = 2.32658\ .
\]

\begin{figure}[htbp]
 % \centering
 % \vspace{-1.7cm}
  \hspace{-1.5cm}
    \epsfig{figure=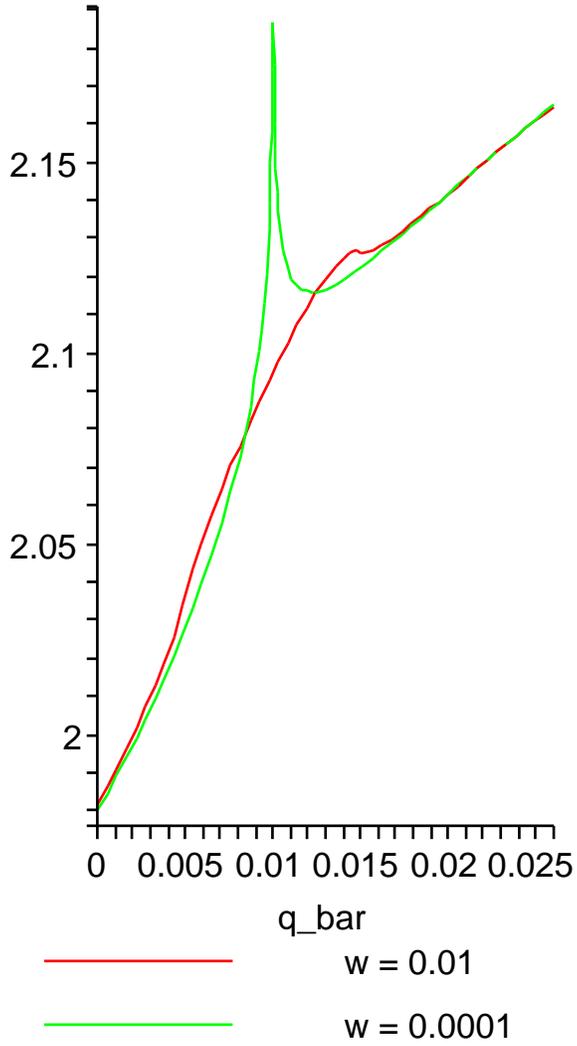}

    \caption[Expected scattering time for box wave packets with constant width
    and varied mean energy] {\it (Color online) The expected scattering time
      $\frac{1}{R_o}\langle \phi |\hat{s}(R_o)| \phi \rangle$ for box wave
      packets is plotted as a function of the dimension-free mean energy
      $\bar{q}$. Plots are shown for packets with constant energy widths $w =
      0.01$ and $w=0.0001$. The mean scattering time has a peak near the
      critical energy at $\rho = 0.01$. The peak is more distinctive when the
      packet is narrower.}
  \label{fig:sbar_2}
\end{figure}

We observe first that the second maximum lies at very high energies. If an
observer is going to send a shell from the radius, say, 1 $m$ to the center,
then the energy needed to achieve the scattering time corresponding to the
second maximum lies at about 25 Earth masses. The existence of the second
maximum and the fall in the scattering time at still higher energies result
from the manipulations of the observer proper time and of his position in the
space-time due to this huge mass concentration rather than from some processes
near and under the horizon. The second observation is that the value of the
second maximum is not much larger than $2R_o$, which corresponds roughly to
the flat space-time value. The plot of the resonance behavior of the time
(\ref{eq:s bar box}) for some typical values of $\rho$ and $w$ in a reasonable
energy interval is given in Fig.\ \ref{fig:sbar_2}.

\section{Discussion}
The values of the scattering times calculated in the previous section are
roughly comparable to the time $2(R_o - R_m)$ the light needs to cross twice
the flat space-time distance between the observer and the mirror. It can be
appreciably shorter if the wave packet contains sufficiently strong
high-energy part ($\bar{q} \approx 1$), or much larger, if it is
concentrated at the resonant energy ($\bar{q} \approx \rho$). Although the
resonance can yield arbitrarily long scattering times, it works only in an
extremely narrow regime of $\bar{q}$ and $w$. Hence, our quantum theory does
not yield black-hole-like objects with their observed properties.

Quite a number of excuses can be thought of. One class of explanations might
be based on the obvious difference between our shell model and a real
astrophysical object. For example, the collapse of the shell is a one-off
event while an astrophysical black hole must be fed steadily in order to be
observable; the zero-rest-mass shell is rather different from a massive star;
etc. Another class contains explanations that seek the reason for too short
scattering times in our calculation method. Could the model of thin null shell
be kept and only the ideas changed of how the scattering time is defined and
calculated? Let us focus on this second possibility.

Some freedom of this class is the usual ambiguity in factor orderings and in
the choice of self-adjoint extensions, such as, in our case, the freedom in
the parameter $\theta$ in Eq.\ (\ref{eq:v theta EV}). It seems, however, that
different choices of this kind are rather unlikely to change the scattering
times a great deal. Some hopeful freedom seems to be the following.

Our quantum calculation is based on the classical formula (\ref{17.1}) that
has been changed in two ways. First, it is extended for energies in the
interval $(0,E_m)$ to the interval $(0,E_o)$ by replacing the parenthesis
under the logarithm by the absolute value signs:
\begin{equation}\label{AV}
  s(R_o) = \sqrt{1-\frac{2GE}{R_o}}\left[2(R_o-R_m)
  +4GE\ln\left|\frac{R_o - 2GE}{R_m - 2GE}\right|\right]\ .
\end{equation}
Second, this extended formula is turned into an operator. Can the formula
(\ref{AV}) be given any ``classical'' meaning for $E > E_m$?

To see how this might be possible, consider first classical shells with
energies smaller than $E_m$. Then the space-time outside the shell can be
foliated by space-like surfaces of constant Schwarzschild time $T$. Let us
denote by $T_m$ the Schwarzschild time of the encounter between the shell and
the mirror, and by $T_\pm$ those between the shell and the observer. One easily
verifies that
\begin{equation}\label{73.1}
  s(R_o) = \sqrt{1-\frac{2GE}{R_o}}\left[2(T_- - T_m)\right]\ .
\end{equation}
Next, this formula can be extended to higher energies as follows. Consider
an in-going shell solution with $E > E_m$. Even then, the Schwarzschild time
coordinate $T$ is well-defined everywhere outside the shell (up to a constant
shift, which has no influence on the results). The differences to the $E <
E_m$ case are that there is the {\em internal} Schwarzschild time for the part
of the space-time that lies inside the horizon, and the {\em external}
Schwarzschild time outside; that the levels of the internal time are time-like;
and that each of these ``times'' runs itself through the whole real axis (in
the maximal---Kruskal---extension of Schwarzschild space-time). Therefore, the
value $T_-$ (of the external) and $T_m$ (of the internal time) are both well
defined in this case, too, and one can again easily verify that Eq.\ 
(\ref{73.1}) gives the same answer as Eq.\ (\ref{AV}).

\begin{figure}[htbp]
 % \centering
 % \vspace{-1.7cm}
  \hspace{4cm}
    \epsfig{figure=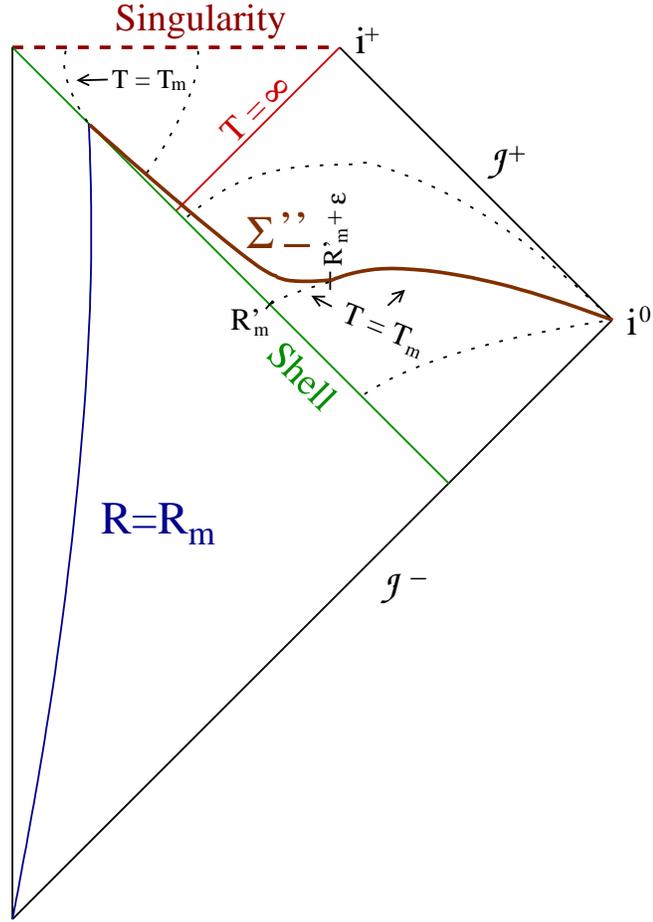}
  
    \caption[In-going shell space-time with the surface $\Sigma_-''$] {\it
      (Color online) The construction of the 3-surface $\Sigma''_-$. The
      dotted lines are the 3-surfaces of constant external and internal
      Schwarzschild time. The space-time above the shell is Schwarzschild and
      below it is Minkowski one. The observer trajectory is not shown, but if
      it crossed the shell at $T_- > T_m$, the scattering time would be
      negative.}
  \label{fig:fold}
\end{figure}

The formula (\ref{73.1}) has an interpretation in terms of space-times and
foliations even for $E > E_m$. Indeed, the external Schwarzschild time
coordinate runs through all real values along the in-going shell, starting by
$-\infty$ at ${\mathcal I}^-$ and reaching $+\infty$ at the horizon, $R =
2GE$. Hence, the external Schwarzschild time takes on the value $T_m$
somewhere at the shell trajectory outside the horizon; let us denote the
radius of this point by $R'_m$; clearly, $R'_m > 2GE > R_m$. Hence, there is a
surface, $\Sigma'$, consisting of two pieces: the first is the shell
trajectory from the radius $R_m$ to the radius $R'_m$ and the second is the
part of the surface $T = T_m$ from $R'_m$ to $R = \infty$. $\Sigma'$ is a
non-time-like surface; it can, therefore, be slightly deformed to a smooth
spherically symmetric {\em space-like} surface $\Sigma_-''$ that runs from the
intersection of the in-going shell with the mirror at $R = R_m$ inside the
horizon until it meets the $T = T_m$ surface outside the horizon at $R = R'_m
+ \epsilon$, where $\epsilon$ is any given real number larger that zero that
can be arbitrarily small. Afterwards, $\Sigma_-''$ coincides with the surface
$T = T_m$ for $R \in (R_m + \epsilon,\infty)$. The construction is displayed
in Fig.\ \ref{fig:fold}.

Suppose that $R'_m < R_o$ so that the scattering time will be positive. Then
the construction of a ``classical'' space-time with a reflected shell and the
scattering time given by Eq.\ (\ref{73.1}) is as follows. Let us cut the
in-going-shell space-time along $\Sigma_-''$, throw away the future part and
denote the remaining space-time by ${\mathcal M}''_-$. Let us further define
another piece ${\mathcal M}''_+$ of space-time as the time-reversal of
${\mathcal M}''_-$ that is given by Eq.\ (\ref{TR}) with an arbitrary choice
of parameter $u$. Finally paste together ${\mathcal M}''_-$ with ${\mathcal
  M}''_+$ along $\Sigma_-''$ and its time reversal $\Sigma_+''$ in ${\mathcal
  M}''_+$ so that spheres of the same radius coincide. The resulting space-time
${\mathcal M}''$ has a continuous metric which is piecewise smooth, but only
$C^0$ at $\Sigma''$ between the radii $R_m$ and $R'_m + \epsilon$ and at the
shell; observe that it is smooth at $\Sigma''$ for $R > R'_m + \epsilon$
because of the time-reversal properties of Schwarzschild space-times mentioned
in the Introduction. The shell is reflected at the mirror that forms a
boundary of ${\mathcal M}''$ and the corresponding scattering time is again
given by Eq.\ (\ref{AV}) for any $E \in (0,E_o)$.

The use of the time reversal in the construction is not accidental. Each
space-time with the outgoing shell that has the same energy and with the
mirror that has the same radius as our in-going-shell space-time is a time
reversal (\ref{TR}) of the in-going-shell space-time. Hence, the space-time
that describes the reflection of the shell at the mirror can also be
constructed by pasting these in-going- and outgoing-shell space-times along
some space-like 3-surfaces that cross the mirror at the shell-reflection
points. One of these surfaces lies in the in-going-, the other in the
outgoing-shell space-time. The surfaces must be isometric to each other or
else they will not match each other and the resulting metric will not be $C^0$
at the pasting points. One can then easily verify that the two 3-surfaces must
also be related by the time reversal.

Let us call ``fold'' any space-like 3-surface where the metric is only $C^0$.
Of course, the folds do not make much sense from the classical general
relativity point of view. A space-time with folds can, however, be used as a
possible path in the calculation of a path integral defined by the Feynman-Kac
formula (Ref.\ \cite{R-S}, Sec.  X.11). Similar space-times have been
described in Refs.\ \cite{H-S-W} and \cite{H}. Observe that the folds in any
polygonal ``zig-zag'' path lie at the surfaces of constant time and their form
is, therefore, influenced by the chosen foliation. The scattering time depends
not only on the number of folds, but also on the foliation, as mentioned in
the Introduction.

Even if one would take this polygonal space-time idea seriously, one could see
at once that it opens new freedom in addition to giving an interpretation of
Eq.\ (\ref{AV}). Why the internal Schwarzschild time $T_m$ of the reflection
is to coincide with the external Schwarzschild time $T''$ at which the two
space-time pieces ${\mathcal M}''_-$ and ${\mathcal M}''_+$ are to be pasted
together for the radii $R > R''$ ($R'' = R'_m + \epsilon$ in the above
construction)?  Clearly, similar constructions can be carried out for {\em
  any} value of $T''$ whatsoever because the external Schwarzschild time runs
through all real values along the shell. This shows explicitly how different
foliations can lead to very different scattering times.

In fact, it seems that the foliation ought to be chosen such that the folds
are limited to a region with as small radius as possible. Recall that the
constructed space-time had a regular (smooth) classical metric outside $R''$.
Sure, for $E > E_m$, there is no classical solution interpolating between the
two asymptotic (in- and outgoing) states of the shell that does not pass at
least through one fold, but the folds could be banished to the inside of the
horizon in this way.  Thus, the resulting quantum geometry could be very
similar to the classical Schwarzschild geometry outside the horizon, while it
had to differ strongly from it inside. At the same time, the condition that
the folds must not protrude too much through the horizon would also lead to
long scattering times (they have to go out of the horizon at least a little
because they have to meet the 3-surfaces of constant external Schwarzschild
time). This follows from the fact that $T''\rightarrow\infty$ along the shell
as $R''$ approaches the horizon.

One possible conclusion from this discussion is that the quantum theory of
gravitational collapse does not entail any natural formula for the scattering
time. More precisely, the differences in scattering times of one and the same
scattering process arising from various possible definitions of scattering
time cannot be attributed only to different choices of factor orderings
(because their order of magnitude is much larger that that of the Planck
constant). This seems to leave some hope that a method (or even a new
principle) exists, which 1) can be better justified than Eq.\ (\ref{AV}) and
2) would lead to considerably longer scattering times. More research is
needed, before a clear understanding can be established.

%\bibliographystyle{alpha}
%\bibliography{bibdiss}

\begin{thebibliography}{99}
\bibitem{HE}S.~W.~Hawking, G.~F.~R.~Ellis, {\it The Large Scale Structure of
    Space-Time}, Cambridge University Press, Cambridge, 1973.
\bibitem{Gidd}S.~W.~Hawking, Commun.\ Math.\ Phys.\ {\bf 87}, 395 (1982);
  T.~Banks, M.~E.~Peskin and L.~Susskind, Nucl.\ Phys.\ {\bf B244}, 125
  (1984).
\bibitem{sach}A.~D.~Sacharov, JETP {\bf 22}, 241 (1966).
\bibitem{hay}S.~A.~Hayward, arXiv: gr-qc/0506126.
\bibitem{HK}P.~H\'{a}j\'{\i}\v{c}ek and C.~Kiefer, Nucl.\ Phys.\ \textbf{B}
  603, 491 (2001).
\bibitem{H}P.~H\'{a}j\'{\i}\v{c}ek, Nucl.\ Phys.\ \textbf{B} 603, 515 (2001).
\bibitem{Honnef}P.~H\'{a}j\'{\i}\v{c}ek, in {\it Quantum Gravity. From Theory
    to Experimental Search}, Eds.~D. Giulini, C.~Kiefer and C.~L\"{a}mmerzahl,
    Springer, Berlin, 2003. ArXive: gr-qc/0204049.
\bibitem{Paris}P.~H\'{a}j\'{\i}\v{c}ek, Nucl.\ Phys.\ B (Proc.\ Suppl.) {\bf
    80} (2000) CD-ROM supplement. ArXive: gr-qc/9903089.
\bibitem{Goldberger}M.~L.~Goldberger and K.~M.~Watson, {\it Collision theory},
  Wiley, New York, 1964.
\bibitem{Martin}P.~A.~Martin, Acta Phys.\ Aust.\ Suppl.\ {\bf 23}, 157 (1981).
\bibitem{Dollard}J.~D.~Dollard, J.\ Math.\ Phys.\ {\bf 5}, 729 (1964).
\bibitem{Bolle:1983xp}D.~Boll\'e, F.~Gesztesy and H.~Grosse, J.\ Math.\ Phys.\
  2{\bf 4}, 1529 (1983).
\bibitem{Kucharrev}K.~V.~Kucha\v{r}, in G.~Kunstatter et al.\ (Eds.), {\it
    Proceedings of the 4th Canadian Conference on General Relativity and
    Relativistic Astrophysics}, World Scientific, Singapore, 1992.
\bibitem{Isham}C.~J.~Isham, in {\it Integrable Systems, Quantum Groups and
    Quantum Field Theories}, Kluver Academic Publishers, London, 1993.
\bibitem{LWF}J.~Louko, B.~Whiting and J.~Friedman, Phys. Rev. D \textbf{57},
  2279 (1998). 
\bibitem{HKou}P.~H\'{a}j\'{\i}\v{c}ek and I.~Kouletsis, Class. Quantum Grav.
  \textbf{19}, 2529 (2002); \textit{ibid.} \textbf{19}, 2551 (2002); {I}.
  Kouletsis and P. H\'{a}j\'{\i}\v{c}ek, \textit{ibid.} \textbf{19}, 2567
  (2002).
\bibitem{ambrus}M.~Ambrus, {\it How long does it take until a quantum system
    reemerges after a gravitational collapse?}, Inauguraldissertation,
  University of Berne, 2004.
\bibitem{Kuchar2} K.~V.~Kucha\v{r}, Phys.\ Rev.\ {\bf D50}, 3961 (1994).
\bibitem{Regge} T.\ Regge and C.\ Teitelboim, Ann.\ Phys.\ {\bf 88}, 286
  (1974). 
\bibitem{R-S}M.~Reed and B.~Simon, {\it Methods of modern mathematical physics
  II. Fourier analysis, self-adjointness}, Acad.\ Press, New York, 1975.
\bibitem{H-S-W}C.~R.~Stephens, G.'t Hooft, and B.~F.~Whiting, Class.\ Quantum
  Grav.\ {\bf 11}, 621 (1994).

\end{thebibliography}

\end{document}